\documentclass[fdp,a4paper,fleqn%
]{w-art}
\usepackage{times,cite,w-thm}
\theoremstyle{plain}

\theoremstyle{definition}

\usepackage[]{graphicx}
\begin{document}
\DOIsuffix{theDOIsuffix} 
\pagespan{1}{} 



\title[ASch Space-Times]{Asymptotically Schr\"odinger Space-Times}


\author[J. Hartong]{Jelle Hartong\inst{1}%
  \footnote{E-mail:~\textsf{hartong@nbi.ku.dk}}}
\address[\inst{1}]{Niels Bohr Institute, Blegdamsvej 17, DK-2100 Copenhagen {\O}, Denmark} 
\author[B. Rollier]{Blaise Rollier\inst{2}\footnote{E-mail:~\textsf{rollier@itp.unibe.ch}}}
\address[\inst{2}]{Albert Einstein Center for Fundamental Physics, Sidlerstrasse 5, 3012 Bern, Switzerland} 
\begin{abstract}
We first review how asymptotically Schr\"odinger space-times arise in a
natural way by performing a TsT transformation on asymptotically AdS
space-times and some of its consequences. We then give a coordinate
independent definition of a pure Schr\"odinger space-time in terms
of an AdS metric and an AdS null Killing vector. Then,
by analogy with the AdS case, we provide a local and coordinate independent definition of a
Schr\"odinger boundary in
terms of a defining function. We use this to
construct the Fefferman--Graham expansions of locally
Schr\"odinger space-times.
\end{abstract}
\maketitle                   





\section{Introduction}
Asymptotically Schr\"odinger space-times form a natural starting
point to try to extend holographic techniques to the realm of
non-relativistic physics. The main reason for this is that
Schr\"odinger space-times can be thought of as a deformation of AdS
spaces and this deformation is mild enough such that certain
properties of the AdS space directly carry over to the Schr\"odinger
case. Clearly it is of great interest to fully take advantage of
this fact in developing non-relativistic holographic techniques. One
close connection between the Schr\"odinger space and AdS is that the former
possesses as a symmetry group the non-relativistic analogue of the
conformal group. This so-called Schr\"odinger group can be embedded into
$SO(d+2,2)$ and possesses an anisotropic scaling symmetry
of the form $D_z: \,t\rightarrow \lambda^zt,\, x\rightarrow \lambda
x$ where $z\neq1$ is called the dynamical exponent and is such that
the time coordinate scales differently than the spatial ones. There
exists a non-relativistic counterpart of a CFT that is based on this
group \cite{Nishida:2007pj} and from a phenomenological point of
view they could be of relevance to the description of cold atoms at
unitarity \cite{Son:2008ye,Balasubramanian:2008dm}. Next we will
restrict ourselves to the special value $z=2$ for which there is an
extra special conformal symmetry and which is such that the metric
admits a global timelike Killing vector field \cite{us09}.

One of the reasons why it is challenging to develop holographic techniques for
Schr\"odinger backgrounds comes from the fact that it is a non-distinguishing space-time \cite{hubeny,us10} that has an intrinsically Galilean like causal structure. The usual boundary proposals are not guaranteed to apply to such space-times. Indeed, one can show that the Penrose construction applied to the
Schr\"odinger metric fails. Moreover,
because on the causal ladder the space-time is only
non-distinguishing boundary definitions such as
the one given by Marolf and Ross \cite{Marolf:2003ik} (see also
\cite{Flores:2010mt}) are not guaranteed to work. It would be
interesting to study if these methods can nevertheless be
successfully applied to Schr\"odinger spaces. Horava and Melby-Thompson proposed an adaptation of the Penrose construction valid for
anisotropic boundaries \cite{Horava:2009vy}. We will adopt a
slightly different point of view (but leading to similar results) and show how a natural notion of a Schr\"odinger
boundary emerges in terms of a defining function once the
Schr\"odinger metric is expressed in terms of an AdS metric and a
null Killing vector field. Our results on the Schr\"odinger boundary appear to have some overlap with \cite{Duval:2012qr}.

We mention that the last section
was not part of the talk presented at the {\it XVII European Workshop on String Theory} in Padova, Italy, (5-9 September, 2011) by B.R. and is added to give a more complete picture.

\section{Asymptotically Schr\"odinger spaces from TsT}
Schr\"odinger solutions can arise in various gravitational theories
but from the holographic point of view solutions embedded into
string theory are the most relevant ones. It is well-known that
$z=2$ (asymptotically) Schr\"odinger solutions can be obtained as
the result of a TsT (T-duality, shift, T-duality) transformation
\cite{Lunin:2005jy} performed on an (asymptotically) AdS space-time.
Such a transformation can be thought of as a solution generating
technique relating different solutions of type II supergravities.
The TsT transformation will produce an asymptotically Schr\"odinger
space-time (ASch) out of a solution of the form AAdS$_5\times Y_5$
where the asymptotically AdS space (AAdS$_5$) is purely gravitational provided that the shift is performed along an isometry of the AAdS$_5$ that
asymptotically becomes null and the T-dualities are done along an isometry of a Sasaki-Einstein internal space $Y_5$. To actually perform the transformation requires adapted
coordinates that make the two isometries involved, forming a 2-torus,
manifest. Let us call $\partial_\varphi$ the isometry on $Y_5$,
$\partial_{\tilde{\varphi}}$ its T-dual and $\partial_V$ the
asymptotically null isometry of the AAdS$_5$ space $g_{\mu\nu}$
along which the shift $V \rightarrow V + \gamma\tilde{\varphi}$ is
performed. Then after a dimensional reduction one finds that out of
the metric $g_{\mu\nu}$ the TsT transformation has produced the set of new fields
\cite{Maldacena:2008wh}
\begin{eqnarray}
\bar{g}_{\mu\nu} & = & e^{-2\Phi/3}\left(g_{\mu\nu}-e^{-2\Phi}A_\mu A_\nu\right)\,, \label{eq: TsT munu}\\
A & = & \gamma e^{2\Phi}g_{V\mu}dx^\mu\,,\\\label{eq: TsT phi}
e^{-2\Phi} & = & 1+\gamma^2g_{VV}\,.
\end{eqnarray}
These fields together solve the equations of motion
coming from the following action
\begin{equation}\label{eq: sch action}
I = \int_{M}d^5x\sqrt{-\bar g}\left(\bar
R-\frac{4}{3}\partial_\mu\Phi\partial^\mu\Phi-V(\Phi)-
\frac{1}{4}e^{-\tfrac{8}{3}\Phi}F_{\mu\nu}F^{\mu\nu}-4A_\mu
A^\mu\right)\,,
\end{equation}
where the potential $V$ is given by
$V(\Phi)=4e^{\tfrac{2}{3}\Phi}\left(e^{2\Phi}-4\right)$. The solution
(\ref{eq: TsT munu})-(\ref{eq: TsT phi}) can always be uplifted 
to a solution of type IIB supergravity\footnote{The action \eqref{eq: sch action} is a simplified version of a three scalar action that
forms a consistent reduction of type IIB supergravity on a squashed 5-sphere \cite{Maldacena:2008wh}.} \cite{Maldacena:2008wh}. Further, it can
be shown that the isometries of $\bar{g}_{\mu\nu}$ are all
those isometries of $g_{\mu\nu}$ that commute with the Killing
vector $\partial_V$. Provided that this vector is (asymptotically)
null it will be guaranteed that the metric $\bar{g}_{\mu\nu}$
(asymptotically) inherits the full set of Schr\"odinger isometries.

As a consequence of these observations a natural class of ASch
space-times emerges. Namely, a class formed by the whole set of
solutions obtained by applying a TsT transformation to those pure
AAdS$_5$ space-times that admit a Killing vector $\partial_V$ that
becomes null asymptotically. It is then possible to identify a
conformal class of Schr\"odinger boundaries that can be thought of
as the TsT image of that subclass of AdS boundaries admitting a null
Killing vector\cite{us11}. Moreover, it can be shown that the
on-shell action (\ref{eq: sch action}) evaluated for the class of
solutions (\ref{eq: TsT munu}) to (\ref{eq: TsT phi}) is TsT
invariant meaning that it reduces to the same value the AdS action
had before performing the transformation. This leads to important
constraints for the construction of counterterms to the action
(\ref{eq: sch action}). Indeed, whatever the complete set of
counterterms for the full space of ASch$_5$ solutions is, upon
substituting the solutions (\ref{eq: TsT munu}) to (\ref{eq: TsT
phi}), the counterterms that contribute divergently must together
equal the known counterterms of the usual AAdS$_5$ action.
Furthermore, it can be seen that not only the on-shell action but
actually also thermodynamic properties of TsT transformed AAdS$_5$
black holes such as entropy, temperature and chemical potentials are
all TsT invariant \cite{us11}.

All in all a lot can be learned by the study of asymptotically
Schr\"odinger space-times produced by the TsT transformation. This
is mainly to be attributed to the knowledge one has about AAdS
spaces such as the structure of the boundary and Fefferman--Graham
expansions which can be put to good use in deriving results for ASch
spaces. Moreover, it is very appealing that these ASch solutions can be uplifted to solutions of type IIB supergravity. Nevertheless, TsT transformations have several drawbacks one would
wish to cure. First of all, the transformation is intrinsically
coordinate (and dimension) dependent as the shift isometry needs to
be manifest. Another severe restriction comes from the fact that
this isometry needs to be global rather than just an asymptotic one.
This already suggests that the set of ASch spaces accessible via TsT
is rather restricted and raises the question: Can we generalize the TsT construction? In order to answer this question we first need to formulate precisely what we mean by a Schr\"odinger boundary. This will be the subject of this proceedings paper in which we apply this to work out the Fefferman--Graham expansions for locally Schr\"odinger space-times. In an upcoming paper \cite{us12} we will then use this to construct Fefferman--Graham expansions of certain asymptotically Schr\"odinger space-times.

\section{From AdS to Schr\"odinger}
To cure some of the obstructions observed previously it is crucial
to first reformulate the result of the TsT transformation in a
coordinate independent manner. For this purpose we will now prove
that a Schr\"odinger space-time $\bar{g}_{\mu\nu}$ can always be
written in the form of a generalized Kerr--Schild metric
\cite{Duval:2008jg}
\begin{equation}\label{eq: shifted metric}
\bar{g}_{\mu\nu} = g_{\mu\nu} - A_\mu A_\nu\,,
\end{equation}
where $g_{\mu\nu}$ is the metric of an AdS space-time and $A^\mu$
any null Killing vector (NKV) field of $g_{\mu\nu}$\footnote{Note
that because $A^2=0$ we do not need to specify which metric has been
used to lower the index on $A^\mu$.}. It then follows that (\ref{eq: shifted
metric}) can be used as a coordinate independent definition for a
pure Schr\"odinger space-time. What underlies this relation is that
the Schr\"odinger algebra consists of all AdS isometries that
commute with a NKV whereas at the same time under the shift only
those isometries that commute with $A^\mu$ are preserved.

\begin{proof}
Let $K^\mu$ be an isometry of $g_{\mu\nu}$. It then follows from
(\ref{eq: shifted metric}) that
\begin{equation}
\mathcal{L}_{K}\bar{g}_{\mu\nu} = -
A_\mu\mathcal{L}_{K}A_\nu - A_\nu\mathcal{L}_{K}A_\mu \,,
\end{equation}
where $\mathcal{L}_{K}$ denotes the Lie derivative along $K^\mu$.
Note that if $[K,A]=0$ then $K^\mu$ is also an isometry of
$\bar{g}_{\mu\nu}$. Next we show that the Schr\"odinger algebra is
formed by the commutator of any choice of AdS NKV. Namely,
$sch(d+3)=\left\{X\in
so(2,2+d)\left.\frac{}{}\right|[A,X]=0\right\}$ where $A$ stands for
any AdS NKV (see also \cite{Duval:2012qr}). We already know that
this is true for certain specific choices of $A_\mu$, so we only
need to prove that it is true for any other choice. An AdS$_{d+3}$
NKV corresponds to a NKV of the embedding space-time that belongs to
$so(2,d+2)$. Therefore any NKV can be written as the $so(2,d+2)$
rotation of some preferred one. This preferred Killing vector can be
chosen to be the one that commutes with the Schr\"odinger algebra.
Hence they all do. It follows that the metric $\bar{g}_{\mu\nu}$ has
at least the Schr\"odinger algebra as its isometry group. Finally,
we know from \cite{SchaferNameki:2009xr} that a metric admitting the
Schr\"odinger algebra as (possibly part of) its isometry group is
either a Schr\"odinger or an AdS space-time. Since in \eqref{eq:
shifted metric} the metric $g_{\mu\nu}$ is AdS and the shift is not
a diffeomorphism we conclude that $\bar g_{\mu\nu}$ must be a
Schr\"odinger space-time.
\end{proof}

We want to take advantage of the Kerr--Schild form of the metric
(\ref{eq: shifted metric}) together with the fact that the AdS
metric $g_{\mu\nu}$ admits a Fefferman--Graham (FG) expansion to construct
an expansion for the Schr\"odinger space. To that end, we first need
to figure out what the most general null Killing vector field on AdS
is when expressed in a FG coordinate system. Therefore, we take a
pure AdS space in FG coordinates
\begin{equation}\label{eq: FG pure ads}
g_{\mu\nu}dx^\mu
dx^\nu=\frac{dr^2}{r^2}+\frac{1}{r^2}\left(g_{(0)ab}+r^2g_{(2)ab}+r^4g_{(4)ab}\right)dx^adx^b\,,
\end{equation}
where\footnote{The case $d=0$ will be presented in \cite{us12}.} for
$d>0$ $g_{(2)ab}$ and $g_{(4)ab}$ are fully given in terms of
$g_{(0)ab}$, a conformally flat representative of the conformal
boundary \cite{Skenderis:1999nb}. We then solve the Killing equations
$\nabla_{(\mu} A_{\nu)} = 0$ in this background and find that the
1-form $A_\mu$ is given by
\begin{eqnarray}
A_r &=& r^{-1}\sigma\,, \label{eq: sol Killing1}\\
A_a &=& g_{ab}A^b_{(0)} - \frac{1}{2}\partial_a\sigma -
\frac{r^2}{4}g_{(2)a}^{\quad \;\,b}\partial_b\sigma \label{eq: sol
Killing2}\,,
\end{eqnarray}
where $\sigma = (d+2)^{-1}\nabla^{(0)}_{a}A^a_{(0)}$ and the
boundary vector field $A^a_{(0)}$ must satisfy the conformal Killing
equation
\begin{equation}
\nabla^{(0)}_{a} A_{(0)b} + \nabla^{(0)}_{b} A_{(0)a} = 2\sigma
g_{(0)ab}\,.
\end{equation}
Here $\nabla^{(0)}$ denotes the covariant derivative with respect to
$g_{(0)ab}$ and $A_{(0)a}=g_{(0)ab}A_{(0)}^b$. We note that for $d>0$
equations (\ref{eq: sol Killing1}) and (\ref{eq: sol Killing2}) form
the most general solution to the Killing equations on a locally AdS space-time and are exact to
all orders. However, we are interested only in null Killing vectors.
We therefore have to additionally impose the extra condition
$A^2=0$. To lowest order this translates into the condition
$g_{(0)ab}A_{(0)}^aA_{(0)}^b=0$ which means that one should only
consider null conformal Killing vectors $A_{(0)}^a$. Further, since any null Killing vector on AdS is automatically hypersurface orthogonal as can be seen by using the Raychaudhuri equation, it follows, from the leading term in
the FG expansion of $A_{[\mu}\nabla_\nu A_{\rho]}=0$, that $A_{(0)}^a$ is also hypersurface orthogonal with respect to the boundary metric $g_{(0)ab}$. It can be shown \cite{us12} that for a conformally flat $g_{(0)ab}$ and a hypersurface orthogonal null conformal Killing vector $A_{(0)}^a$ the higher order terms in the condition $A^2=0$ are automatically satisfied and thus do not impose any further restrictions on $A_{(0)}^a$.

\section{Locally Schr\"odinger space-times and the Schr\"odinger boundary}
As already mentioned in the introduction this section was not part
of the presentation given at the \textit{XVII European Workshop on
String Theory 2011} in Padova. We will now make our way towards a
definition of locally Schr\"odinger space-times by working by
analogy with locally AdS spaces. In doing so, we will obtain a
definition for the Schr\"odinger boundary in terms of a defining
function. We start by computing the Riemann tensor for a pure
Schr\"odinger space from (\ref{eq: shifted metric})
\begin{equation}
\bar{R}_{\mu\nu\rho\sigma} = -
\bar{g}_{\mu\rho}\bar{g}_{\nu\sigma}+\bar{g}_{\mu\sigma}\bar{g}_{\nu\rho}
+ \bar{g}_{\mu\rho}A_\nu A_\sigma - \bar{g}_{\nu\rho}A_\mu A_\sigma
+ \bar{g}_{\nu\sigma}A_\mu A_\rho-\bar{g}_{\mu\sigma}A_\nu A_\rho
+\frac{3}{4}F_{\mu\nu}F_{\rho\sigma}\label{eq: Riemann tensor Sch}\,,
\end{equation}
using that $A^\mu$ is a hypersurface orthogonal null Killing vector\footnote{Note that
$A^\mu$ is actually a hypersurface orthogonal null Killing vector of $g_{\mu\nu}$ as well as of
$\bar{g}_{\mu\nu}$.} and denoting $F_{\mu\nu}=\partial_\mu A_\nu -
\partial_\nu A_\mu$. We use this result to introduce the notion of a
locally Schr\"odinger space-time, in analogy with locally AdS
space-times, as those metrics $\bar{g}_{\mu\nu}$ satisfying
(\ref{eq: Riemann tensor Sch}) as well as admitting (locally) a null
Killing vector $A^\mu$. A straightforward property of the
Schr\"odinger Riemann tensor we will make use of is that upon
contraction with $A^\mu$ it can be seen to behave as an AdS Riemann
tensor
\begin{equation}\label{eq: Contracted Riemann}
\bar{R}_{\mu\nu\rho\sigma}A^\sigma =
(-\bar{g}_{\mu\rho}\bar{g}_{\nu\sigma} +
\bar{g}_{\mu\sigma}\bar{g}_{\nu\rho})A^\sigma\,.
\end{equation}

With the use of (\ref{eq: shifted metric}) and (\ref{eq: Contracted
Riemann}) we define a Schr\"odinger boundary in terms of a defining
function $\Omega(r,x)$. First, let $\Omega=0$ denote the location of
the Schr\"odinger boundary. We will consider a conformal rescaling
of both the AdS and the Schr\"odinger metric. Denoting by a tilde
the conformally rescaled metrics we have
$\tilde{g}_{\mu\nu}=\Omega^2g_{\mu\nu}$ and
$\tilde{\bar{g}}_{\mu\nu} = \Omega^2\bar{g}_{\mu\nu}$. Under such a
conformal rescaling equation (\ref{eq: shifted metric}) transforms
as
\begin{equation}\label{eq: conf rescaled shifted metric}
\tilde{\bar{g}}_{\mu\nu} = \tilde{g}_{\mu\nu} -
\Omega^{-2}\tilde{A}_\mu \tilde{A}_\nu\,,
\end{equation}
where $\tilde{A}_\mu=\tilde{g}_{\mu\nu}A^\nu$. Following the same
procedure as in \cite{Graham:1999jg} we rewrite the Schr\"odinger
Riemann tensor in terms of the conformally rescaled quantities. To
leading order in a near boundary expansion we find
\begin{equation}\label{eq: Riem near boundary exp}
\bar{R}_{\mu\nu\rho\sigma} = \Omega^{-4} ( -
\tilde{\bar{g}}_{\mu\rho}\tilde{\bar{g}}_{\nu\sigma} +
\tilde{\bar{g}}_{\mu\sigma}\tilde{\bar{g}}_{\nu\rho})\tilde{\bar{g}}^{\kappa\tau}\partial_\kappa\Omega\partial_\tau\Omega
+ \mathcal{O}(\Omega^{-3})\,.
\end{equation}
It then directly follows from (\ref{eq: Contracted Riemann}) and the
contraction of (\ref{eq: Riem near boundary exp}) with $A^\sigma$
that
\begin{equation}\label{eq: cond Omega}
\left.\Omega^{-2}\bar{g}^{\kappa\tau}\partial_\kappa\Omega\partial_\tau\Omega\right|_{\Omega=0}
= 1\,.
\end{equation}
On the other hand, since $\Omega^{-2}\tilde{g}_{\mu\nu}$ is an AdS
metric, $\Omega$ must also satisfy the properties of an AdS defining
function implying for similar reasons as above that
$\left.\Omega^{-2}g^{\kappa\tau}\partial_\kappa\Omega\partial_\tau\Omega\right|_{\Omega=0}
= 1$. Hence, by substituting the expression for the inverse
Schr\"odinger metric $\bar{g}^{\mu\nu} = g^{\mu\nu} + A^\mu A^\nu$
into (\ref{eq: cond Omega}) we learn that
\begin{equation}\label{eq: cond Omega 2}
\left.\Omega^{-2}(A^\mu\partial_\mu\Omega)^2\right|_{\Omega=0} =
0\,.
\end{equation}
As $\partial_\mu\Omega$ is normal to the boundary we
arrive at the important conclusion that $A^\mu$ must necessarily be
tangential to the Schr\"odinger boundary\footnote{We mention that the so-called lightlike
lines on the Schr\"odinger space are geodesics whose tangent is
precisely $A^\mu$. In \cite{us10} we showed that these lightlike
lines play a key role in describing the non-relativistic causal
structure of the Schr\"odinger space-time. The fact they are
tangential to the boundary is strongly indicative of the boundary
inheriting this non-relativistic causal structure. }.

Let us now go back to the AdS space point of view. The conformal
rescaling $\tilde{g}_{\mu\nu} = \Omega^2g_{\mu\nu}$ transforms the
Killing equation for the null Killing vector $A^\mu$ into
\begin{equation}\label{eq: conf rescaled Killing eq}
\tilde{\nabla}_\mu\tilde{A}_\nu + \tilde{\nabla}_\nu\tilde{A}_\mu =
2\tilde{g}_{\mu\nu}A^\rho\partial_\rho \log\Omega\,,
\end{equation}
where $\tilde A_\mu=\tilde g_{\mu\rho}A^\rho$. The projection of (\ref{eq: conf rescaled Killing eq}) onto the
boundary is
\begin{equation}\label{eq: projected conf rescaled Killing eq}
\tilde{h}_\mu{}^{\rho}\tilde{h}_\nu{}^{\sigma}\tilde{\nabla}_\rho(\tilde{h}_\sigma{}^{\tau}\tilde{A}_\tau)
+
\tilde{h}_\nu{}^{\rho}\tilde{h}_\mu{}^{\sigma}\tilde{\nabla}_\rho(\tilde{h}_\sigma{}^{\tau}\tilde{A}_\tau)
= 2\tilde{h}_{\mu\nu}A^\rho\partial_\rho \log\Omega -
2A^\tau\tilde{n}_\tau \tilde{K}_{\mu\nu} \,,
\end{equation}
where $\tilde{n}^\mu$ denotes the unit normal to the AdS boundary
with respect to the rescaled metric $\tilde{g}_{\mu\nu}$, the
boundary projector $\tilde h_\rho{}^\mu$ is given by $\tilde{h}_\rho{}^{\mu} =
\delta_\rho^{\mu}- \tilde{n}_\rho \tilde{n}^\mu$ and
$\tilde{K}_{\mu\nu}=\tilde{h}_\mu{}^{\sigma}\tilde{\nabla}_{\sigma}\tilde{n}_\nu$
is the extrinsic curvature. Evaluating (\ref{eq: projected conf
rescaled Killing eq}) at $\Omega=0$ using (\ref{eq: cond Omega 2})
and the fact that $\tilde{n}_\mu=\partial_\mu\Omega$ near $\Omega=0$
we conclude that the boundary projected vector
$\tilde{h}_\mu{}^{\tau}\tilde A_\tau$ is an AdS boundary Killing vector. It
is straightforward to show that it is also null with respect to the
AdS boundary metric.

As a result we observe that the condition (\ref{eq: cond Omega 2})
restricts the AdS boundary metrics to those admitting a
null Killing vector. This condition restricts the expansion given by
(\ref{eq: sol Killing1}) and (\ref{eq: sol Killing2}). In other
words we find that the AdS decomposition of the Schr\"odinger metric
(\ref{eq: shifted metric}) relates the Schr\"odinger boundary to
those AdS boundaries whose metric admits a null Killing vector. In
the previous section starting from (\ref{eq: shifted metric}) we
could have used any AdS Fefferman--Graham coordinate system to
perform the shift to Schr\"odinger. This is certainly possible but
in this way we would also consider coordinates such as global AdS
coordinates and write the Schr\"odinger metric in terms of those.
Without any further restrictions this approach gives expressions for
the Schr\"odinger metric that appear to be in Fefferman--Graham form
whereas they are really not unless we had imposed from the start to
only take into account those AdS coordinates satisfying (\ref{eq:
cond Omega 2}). For example had we chosen global AdS coordinates we
would end up with an expression for the Schr\"odinger metric in
which the Schr\"odinger boundary is no longer manifest because
$A^\mu$ is not tangent to the global AdS boundary.

We finally proceed to the construction of a Fefferman--Graham
expansion for locally Schr\"odinger space-times based on these new
observations. Using (\ref{eq: shifted metric}), the well-known
Fefferman--Graham expansions for locally AdS space-times
\cite{Skenderis:1999nb} and the results of the previous section
together with the new restrictions on the AdS boundary metric to
admit a null Killing vector we obtain for $d>0$
\begin{eqnarray}
d\bar{s}^2 &=& \frac{dr^2}{r^2} + \left( \frac{\tilde{h}_{ab}}{r^2}
- \frac{\tilde{A}_a\tilde{A}_b}{r^4}\right)dx^adx^b \,,\\
\tilde{h}_{ab} &=& g_{(0)ab} + r^2g_{(2)ab} + \frac{r^4}{4}g^{\quad\;\; c}_{(2)a}g_{(2)cb}\\
A_a &=& r^{-2}\tilde{A}_a = r^{-2}\tilde{h}_{ab}A^b_{(0)} \,,\\
A_r &=& 0\,,
\end{eqnarray}
in which $g_{(0)ab}$ is a conformally flat representative of the
conformal boundary that admits $A^a_{(0)}$ as a 
null Killing vector and $g_{(2)ab}$ is given by
\begin{equation}
g_{(2)ab} = -\frac{1}{d}\left(R_{(0)ab}-
\frac{R_{(0)}}{2(d+1)}g_{(0)ab}\right)\,.
\end{equation}

\begin{acknowledgement}
We express our gratitude to Matthias Blau for many useful
discussions. The work of J.H. was supported in part by the Danish National Research Foundation 
project ``Black holes and their role in quantum gravity''. The work of B.R. (and in early stages of the project also the work of J.H.) was supported in part by the Swiss National Science Foundation and the ``Innovations- und Kooperationsprojekt
C-13'' of the Schweizerische Universit\"atskonferenz SUK/CRUS.
\end{acknowledgement}

\end{document}